\documentclass[aps,pra,groupedaddress,nofootinbib,reprint]{revtex4-2}

\usepackage{amsmath}
\usepackage[colorlinks=true,citecolor=black,linkcolor=black]{hyperref}
\usepackage{amsfonts}
\usepackage{amssymb}
\usepackage{mathrsfs}
\usepackage{mathtools}

\usepackage{tikz}

\newcommand\copyrighttext{%
  \footnotesize \copyright~2025. This manuscript version is made available
  under the CC BY-NC-ND 4.0 license
  \href{https://creativecommons.org/licenses/by-nc-nd/4.0/}
  {https://creativecommons.org/licenses/by-nc-nd/4.0/}.}
\newcommand\copyrightnotice{%
\begin{tikzpicture}[remember picture,overlay]
\node[anchor=south,yshift=10pt] at (current page.south) {\parbox{\dimexpr\textwidth-\fboxsep-\fboxrule\relax}{\copyrighttext}};
\end{tikzpicture}%
}

\usepackage[T1]{fontenc}
\usepackage{times}
\usepackage{fix-cm}

\usepackage{natbib}
\usepackage{graphicx}
\usepackage[labelfont=footnotesize,font=footnotesize]{caption} 
\usepackage[labelfont=scriptsize,font=scriptsize,singlelinecheck=off,justification=centering]{subcaption}
\usepackage{booktabs} 
\usepackage{tabularx}
\usepackage{tabulary}
\newcolumntype{Z}{>{\centering\arraybackslash}X} 

\newcommand{\ket}[1]{\mathop{\left|#1\right\rangle}}

\newcommand{\bpm}{\begin{pmatrix}}
\newcommand{\epm}{\end{pmatrix}}

\newcommand{\hilb}{\mathscr{H}}

\newcommand{\mink}{\EE^{1, 3}}

\newcommand{\Spin}{\mathrm{Spin}}

\newcommand{\SUTWO}{\mathrm{SU}(2)}

\newcommand{\SOTHREE}{\mathrm{SO}(3)}

\newcommand{\dd}[1]{\mathrm{d}#1}

\newcommand*\diff{\mathop{}\!\mathrm{d}}
\newcommand{\ppp}{\mathbf{p}}

\newcommand{\CC}{\mathbb{C}}
\newcommand{\EE}{\mathbb{E}}

\newcommand{\VC}{\mathscr{V}}

\begin{document}

\title{Proposal for an experiment to verify Wigner's rotation at
non-relativistic speeds with massive spin-$1/2$ particles}
\author{Veiko Palge}
\email[Email: ]{veiko.palge@ut.ee}
\affiliation{Laboratory of Theoretical Physics, Institute of Physics,
University of Tartu, W. Ostwaldi 1, 50411 Tartu, Estonia}

\author{Jacob Dunningham}
\email[Email: ]{J.Dunningham@sussex.ac.uk}
\affiliation{Department of Physics and Astronomy, University of Sussex,
Brighton BN1 9QH, United Kingdom}

\author{Yuji Hasegawa}
\email[Email: ]{hasegawa@ati.ac.at}
\affiliation{Atominstitut, Vienna University of Technology, Stadionalle 2, 1020
Vienna, Austria}

\author{Christian Pfeifer}
\email[Email: ]{christian.pfeifer@zarm.uni-bremen.de}
\affiliation{Center of Applied Space Technology and Microgravity (ZARM),
University of Bremen, Am Fallturm 2, 28359 Bremen, Germany}

\begin{abstract}
The Wigner rotation of quantum particles with spin is one of the fascinating
consequences of interplay between special relativity and quantum mechanics. In
this paper we show that a direct experimental verification of Wigner's rotation
is in principle accessible in the regime of \emph{non-relativistic} velocities
at $\,\sim\!\!10^3$~$\mathrm{m s^{-1}}$ for massive spin-$1/2$ particles. We
discuss how the experiment could be carried out in a laboratory using cold
neutrons. The measurement at non-relativistic velocities becomes possible
through letting neutrons propagate for a sufficiently long time because Wigner
rotation is a cumulative effect.
\end{abstract}

\maketitle
\copyrightnotice

\section{Introduction}

Over the years, low-energy neutrons have proven to be invaluable in a range of
experiments that probe foundational issues in physics \cite{rauch_werner_2000,
sponar_etal_2010, hasegawa_etal_2006, bartosik_etal_2009, klepp_etal_2014,
sponar_2021}. They are useful test particles in interferometric, spectroscopic
and scattering experiments probing low-energy scales and are a complementary
tool to particle colliders. Their significance lies in providing direct tests
of fundamental concepts to a high degree of accuracy. One area they might be
usefully deployed is in precision measurements of the Wigner rotation, which
first came to the fore in the context of atomic physics when Thomas showed that
the hydrogen atom needs a relativistic correction due to a phenomenon that came
to be called the Thomas precession \cite{thomas_motion_1926}. In a nutshell,
the Wigner rotation is a phenomenon that if a physical system undergoes
non-collinear Lorentz boosts, then it is rotated relative to the original
frame. Wigner's rotation also holds in the quantum domain and is experimentally
confirmed in a plethora of phenomena from atomic spectra to particle
acceleration. In the context of quantum information, it is the reason why the
behavior of entanglement in relativity differs significantly from the standard,
non-relativistic theory \cite{gingrich_quantum_2002, peres_quantum_2002,
friis_relativistic_2010, palge_behavior_2015, barr_etal_2023}.

In this paper we propose an experiment to verify Wigner's rotation in the
regime of what are typically regarded as non-relativistic velocities at
$\,\sim\!\!10^3$~$\mathrm{m s^{-1}}$ in an implementation with cold neutrons.
This is significant because it would provide \emph{direct} evidence of Wigner's
rotation in the quantum domain in a system that can be carefully controlled
and, because the experiment does not require relativistic speeds, it can be
carried out in a laboratory. From a geometric point of view the experiment can
be regarded as explicitly verifying the curvature of the relativistic velocity
space---a feature that sets apart the relativistic and non-relativistic
theories of spacetime \cite{palge_pfeifer_2024}.

The article is organized as follows. We begin by summarizing the theoretical
derivation of the Wigner rotation in Sec.~\ref{sec:wigrot} followed by the
discussion of the experimental setup with cold neutrons in Sec.~\ref{sec:exp}
and the conclusion in Sec.~\ref{sec:conclusion}.

\section{Wigner rotation}\label{sec:wigrot}

Wigner's rotation originates in the fact that the subset of pure Lorentz boosts
does not form a subgroup of the Lorentz group \cite{halpern_special_1968,
costella_2001, rhodes_relativistic_2004}. The combination of two boosts
$\Lambda(v_1)$ and $\Lambda(v_2)$ is in general a boost \emph{and} a rotation,
\begin{align}
\Lambda(v_2) \Lambda(v_1) = W[\alpha(v_1, v_2)] \Lambda(v_3),
\end{align}
where, for massive spin-$1/2$ particles that we will focus on, $W(\alpha) \in
\mathrm{SO}(3)$ is the Wigner rotation with angle $\alpha$ about the axis
$\hat{n} = \hat{v}_2 \times \hat{v}_1$ orthogonal to the plane specified by
$v_1$ and $v_2$. This effect is a truly relativistic phenomenon both in the
conceptual and quantitative senses. The latter means that in a simple scenario
consisting of two boosts the rotation angle becomes significant at large
velocities. For instance, if $v_{1, 2} = 0.5 c$ and the velocities are
orthogonal, then the rotation is $8$ deg, while for non-relativistic speeds
\mbox{$v_{1, 2} = 2 \cdot 10^3\,$}~$\mathrm{m s^{-1}}$ it is barely measurable
at $\,\sim\!\! 10^{-5}$~deg. However, one can achieve a measurable effect at
low, non-relativistic velocities by noting another property of the Wigner
rotation, namely, that it is \emph{cumulative}. This is because Wigner's
rotation is ultimately a \emph{holonomy}---a geometric phase that accumulates
when a particle follows a path in the momentum space, in analogy to how Berry's
phase accumulates when a particle follows a path in the parameter space
associated with the Hamiltonian \cite{wilczek_shapere_1989, lyre_2014,
palge_pfeifer_2024}.

In the following, we will take the geometric approach
\cite{palge_pfeifer_2024}, which is based on the fiber bundle theory
\cite{simon_1983_holonomy, cisowski_2022}, and give an outline of how the
Wigner angle can be calculated based on the idea that it is a geometric phase.
We will then use the result to design an experiment that allows to measure the
Wigner rotation of a particle at non-relativistic speeds.

From the intuitive point of view, in the fiber bundle picture we can think of
the state of the massive relativistic spin-$1/2$ particle as a field on
\emph{curved} momentum space. When a particle is Lorentz boosted it follows a
geodesic in this manifold. This means the spin of the particle is parallel
transported along the geodesic. Because the momentum space is curved, the state
of the particle in general changes non-trivially when it is parallel
transported along a given path. Hence, the final state after completing
parallel transport along a given path is generally not identical to the initial
state. For closed paths, the transformation that maps the initial state to the
final state is called a holonomy. The Wigner rotation is thus nothing but the
holonomy which arises due to the curvature of the relativistic momentum space.
We will next describe how to compute the holonomy transformation for a massive
relativistic spin-$1/2$ particle.

In the fiber bundle theory, the quantum relativistic massive spin-$1/2$
particle is described by the spinor bundle \mbox{$(\pi: E \rightarrow \VC^+_m,
\hilb_\ppp)$} where, using the sign convention $\eta\sim(+,-,-,-)$, $\VC^+_m$
is the forward mass hyperboloid $\VC^+_{m,x} = \{ P \in T^*_x \mink \; | \;
\eta(P, P) = \eta^{\mu\nu} p_\mu p_\nu = m^2, p_0 > 0 \} \subset  T^*_x
\mink\,$ and the typical fiber $\hilb_\ppp = \CC^2$, for a more detailed
account see \cite{palge_pfeifer_2024}. Because the Minkowski space is flat, we
can identify the mass hyperboloids at different points of spacetime and speak
of a single momentum hyperboloid $\VC^+_m$ in which the particle moves. Using
the spherical polar coordinates, the intrinsic geometry of the momentum
hyperboloid is described by the metric
\begin{align}\label{eq:Hmet}
g = -\left( \tfrac{m^2}{E^2}     \dd \rho  \otimes \dd \rho
  + \rho^2               \dd\theta \otimes \dd\theta
  + \rho^2 \sin^2 \theta \dd\phi \otimes \dd\phi\right) \, .
\end{align}
In order to describe parallel transport, we compute the coefficients
$\omega^A{}_B$ of the spin connection using the relation
\begin{align}\label{eq:connection-general-1}
\omega^A{}_{B i} = e^A{}_k e_B{}^j{} \Gamma^k{}_{ij}
+ e^A{}_k \partial_i e_B{}^k{}\,,
\end{align}
where $e^A{}_i$ are components of an orthonormal coframe \mbox{$\Theta^A =
e^A{}_i \diff z^i$} of the momentum space metric~$g$. We can display $\omega$
as the matrix
\begin{align}\label{eq:local-connection-form-mass-shell-1}
\omega &=
\begin{bmatrix}
0    &   -\frac{E}{m} \diff\theta &    -\sin\theta\, \frac{E}{m} \diff\phi     \\
\frac{E}{m} \diff\theta     &    0     &        -\cos\theta \diff\phi  \\
\sin\theta\, \frac{E}{m} \diff\phi     &   \cos\theta \diff\phi    &     0
\end{bmatrix}.
\end{align}

We can likewise express the curvature of the Levi-Civita connection as a
collection of $2$-forms $\Omega^A{}_B$ over $\VC^+_{m}$ written as the matrix
\begin{widetext}
\begin{align}\label{eq:Lso3-curvature-1}
\Omega = &
\begin{bmatrix}
0 & - \frac{\sqrt{E^2-m^2}}{E m} \diff\rho \wedge \diff\theta & - \frac{\sqrt{E^2-m^2}}{E m} \sin\theta \diff\rho \wedge \diff\phi \\
\frac{\sqrt{E^2-m^2}}{E m} \diff\rho \wedge \diff\theta &  0 & -\frac{(E^2-m^2)}{ m^2}  \sin\theta \diff\theta \wedge \diff\phi \\
\frac{\sqrt{E^2-m^2}}{E m} \sin\theta \diff\rho \wedge \diff\phi & \frac{(E^2-m^2)}{ m^2}  \sin\theta \diff\theta \wedge \diff\phi  &  0
\end{bmatrix}.
\end{align}
\end{widetext}

The state space of the particle is given as the space of square integrable
sections $\psi: \VC^+_m \to \mathbb{C}^2$ of the spinor bundle with a suitable
representation of the group $\SUTWO$. Formally, the spinor bundle is the
associated bundle to the principal spin bundle whose structure group $\Spin(3)
\cong \SUTWO$ is the double cover of $\SOTHREE$.

In order to describe parallel transport of spinors, we need the spinor
connection. This is induced by the Levi-Civita connection via lifting the
$\SOTHREE$ connection to the spin connection, and then generating the spinor
connection. We get
\begin{align}\label{eq:connSpin}
\omega_s
= - \frac{i}{2} \left( \frac{E}{m}            \diff\theta \, \sigma^3
                        - \frac{E}{m} \sin\theta \diff\phi   \, \sigma^2
                        +  \cos\theta            \diff\phi   \, \sigma^1
		\right)\,.
\end{align}
Using the spinor connection, we calculate the spinor curvature
\begin{align}\label{eq:spinor-curvature-1}
\Omega_s = \diff\omega_s + \omega_s \wedge \omega_s,
\end{align}
which yields
\begin{align}\label{eq:SinC}
\Omega_s =
&\frac{i}{2} \left(- \frac{\sqrt{E^2-m^2}}{E m} \diff\rho \wedge \diff\theta \,
\sigma_3 \right. \nonumber \\
&\phantom{AA} + \frac{\sqrt{E^2-m^2}}{E m} \sin\theta \diff\rho \wedge
\diff\phi\, \sigma_2 \nonumber \\
&\phantom{AA} - \left. \frac{E^2-m^2}{m^2} \sin\theta \diff\theta \wedge
\diff\phi \,     \sigma_1
\right).
\end{align}

Consider now a particle that follows circular path $C$ in the momentum space.
This means the particle is moving with constant speed $v$, but constantly
changing its direction. One can think of particle's movement as undergoing
infinitesimal parallel transports when it travels around the circular
trajectory. Each small parallel transport corresponds to a small boost which
gives rise to Wigner's rotation, all of which accumulate when the particle has
completed one orbital revolution. This corresponds to the famous case of Thomas
precession, which is formally a holonomy associated with a circular path in the
momentum space. The holonomy matrix can be calculated in terms of the spinor
curvature $\Omega_s$
\begin{align}\label{eq:holonomy-matrix-via-curvature-1}
\mathrm{Hol}\left( \omega_s, C \right) = \exp\left( - \int_D \Omega_s \right),
\end{align}
where $D$ is a disk with boundary $C$ and the matrix is a function of
connection $\omega_s$ and path $C$. In our case, the path is a circle with
radius $\rho_0 \in  \VC^+_m$, where $\rho_0 = m v/\sqrt{1-v^2}$. It is the
boundary of the disc $D$ given by $\theta = \pi/2$, $\rho \in (0,\rho_0)$ and
$\phi\in (0,2\pi)$. Using \eqref{eq:SinC} with $\diff\theta = 0$, the integral
\begin{align}
- \int_D \Omega_s
&= - \int_0^{\rho_0} \int_{0}^{2\pi} \diff\rho \diff\phi\ \frac{i}{2} \frac{\rho}{\sqrt{m^2+\rho^2}m} \ \sigma_2\\
&= - i \sigma_2  \pi  \left( \gamma(v) - 1 \right)\,.
\end{align}
By substituting into \eqref{eq:holonomy-matrix-via-curvature-1} we obtain the
matrix
\begin{align}\label{eq:ThoPrecFin}
\mathrm{Hol}(\alpha) = \exp\left(-i \frac{\alpha}{2} \sigma_2\right) =
  \begin{pmatrix}
	\cos(\alpha/2) & -\sin(\alpha/2) \\
   \sin(\alpha/2) & \cos(\alpha/2)
\end{pmatrix}\,,
\end{align}
where we use the argument $\alpha$ to emphasize that this is an $\SUTWO$
rotation matrix with an angle
\begin{align}\label{eq:holonomy_angle}
\alpha(v) = 2\pi \left( \gamma(v)- 1 \right)
\end{align}
which represents the holonomy transformation of the spin state after the
particle has completed one revolution around~$C$.

We will next describe an experiment that allows to measure Wigner's rotation
using cold neutrons.

\section{Experiment}\label{sec:exp}

\begin{figure}[hbt]
\centering
\includegraphics[width=0.4\textwidth]{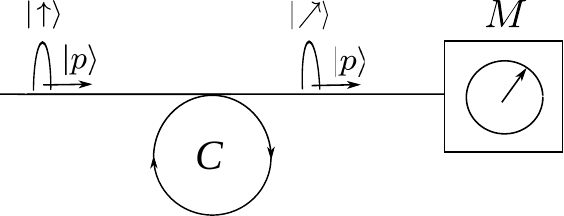}
\caption{\label{fig:scheme_1_4}%
Experimental setup for measuring Wigner rotation. Path of the particle is shown
as solid line. The particle with momentum $\ket{p}$ and spin $\ket{\uparrow}$
enters from the left and follows circular path $C$. After each orbital
revolution the spin is transformed by $\mathrm{Hol}(\alpha)$ resulting in
rotation by angle $\alpha$. Letting the particle orbit $N$ times produces the
spin rotation by angle $N \alpha$. After this the particle in the rotated spin
state $\ket{\nearrow}$ exits towards the measurement apparatus $M$ which
measures the direction of spin.}
\end{figure}

Consider a setup shown in Fig.~\ref{fig:scheme_1_4} where a spin-$1/2$ particle
with mass $m$ and momentum $\ket{p}$ comes in from the left with the initial
spin prepared in state $\ket{\uparrow}$. We assume that the momentum and spin
lie in the plane of the figure.
The particle then follows a path around circle $C$ where it undergoes a series
of small non-collinear boosts, each of which produces a tiny Wigner rotation.
The total Wigner angle accumulated as a result of a single orbital revolution
around the circle at constant speed $v$ is given by \eqref{eq:holonomy_angle}.
While this angle is still too small for detection in practice, the effect can
be amplified by letting the particle travel around the circle $N$ times. This
means the initial $\ket{\uparrow}$ and final spin state $\ket{\nearrow}$ of the
particle are related by applying the holonomy transformation $N$ times
\begin{align}\label{eq:finalinitialspin}
\ket{\nearrow} = \mathrm{Hol}(\alpha)^N \ket{\uparrow} = \mathrm{Hol}(N \alpha)
\ket{\uparrow},
\end{align}
thus producing the total rotation angle $N \alpha$.\footnote{Note that the last
equality in \eqref{eq:finalinitialspin} is simply obtained by using the
explicit form of the holonomy matrix \eqref{eq:ThoPrecFin} for the case under
consideration.}

A measurement $M$ carried out on the final spin state shows the difference
between the directions of the initial and final spins. Note that since we
assumed that the initial spin lies in the plane of the circle or, equivalently,
in the plane of the boosts around the circle, the particle undergoes maximal
Wigner rotation.

\subsection{Cold neutrons}

Cold neutrons provide an interesting platform for realizing this experiment.
They have been used extensively to probe fundamental concepts of quantum
physics \cite{hasegawa_noncommuting_1999, sponar_violation_2010,
demirel_measurement_2015, sponar_2021}. The wavelength of cold neutrons ($\sim
10^{-9} - 10^{-8}$~m) is larger than the distance between atoms and they mainly
interact with the optical or Fermi pseudo potential $V_F$ of matter
\cite{sponar_2021}. We will therefore first analyze the magnitude of the effect
using a model where neutrons are described in terms of neutron optics. Later in
section \ref{sec:wave_mechanics} we will give a sketch of how this analysis
could be extended to a more detailed analysis using wave mechanics.

We assume a setup similar to the one in Fig.~\ref{fig:scheme_1_4} which
involves cold neutrons traveling at \mbox{$v \sim 10^3$}~$\mathrm{m s^{-1}}$
with the mean lifetime $\tau_{n} = 880$~s or about $15$~minutes. Suppose a
ring-like guide with radius $r$ can be manufactured which makes neutrons follow
a circular path $C$. According to neutron optics, this can be achieved by using
material whose Fermi potential $V_F$ is such that if neutrons approach the
surface of the guide at grazing incidence with small angle $\theta$, they are
reflected by the potential of the material. Reflection occurs if the kinetic
energy $E$ corresponding to the normal component of the velocity is smaller
than the Fermi potential of the material. In other words, for total reflection
$E$ must satisfy the Fermi-Zinn condition $E \sin^2\theta < V_F$. For instance,
for a typical material Ni with $V_F = 250$~neV, the angle $\theta < 0.3$~deg
for neutrons at $v = 10^3$~$\mathrm{m s^{-1}}$. It is important to note that,
because the Fermi potential is spin independent, the reflection does not change
the spin state \cite{boothroyd_2020}.

This means that in the neutron optics approach the circular path is
approximated by a polygon. The number of vertices of the polygon depends on the
angle at which the particle approaches the surface: smaller angle results in
higher number of vertices and vice versa. As the number of vertices grows, the
polygon approaches a circle.

\begin{figure}
\centering
\includegraphics[width=0.45\textwidth]{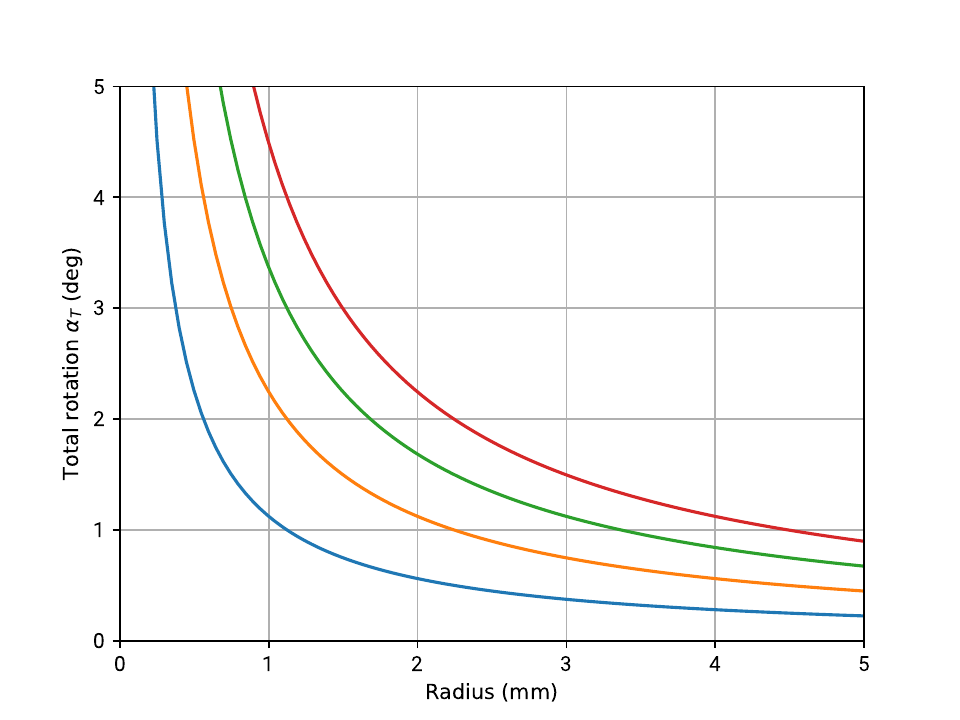}
\caption{\label{fig:wigner_rotation_circle}
Dependence of the total Wigner rotation $\alpha_T$ on the radius $r$ of circle
for a particle with $v = 10^3\,$$\mathrm{m s^{-1}}$, and duration of experiment
$t$ is $4\tau_{n}$ (blue), $8 \tau_{n}$ (yellow), $12 \tau_{n}$ (green) and $16
\tau_{n}$ (red).}
\end{figure}

Assuming the circular neutron guide has radius $r$, the particle makes $v t /
2\pi r$ revolutions in time $t$. Multiplying this with the angle $\alpha(v)$
per one revolution in \eqref{eq:holonomy_angle}, we obtain the total rotation
\begin{align}\label{eq:totalRotation}
\alpha_T = \frac{v t}{2 \pi r} \cdot \alpha(v).
\end{align}
Plot of $\alpha_T$ for cold neutrons with $t = 4\tau_{n}$ as well as longer
times is shown in Fig.~\ref{fig:wigner_rotation_circle}.
This shows different parameter regimes are possible. For instance, a total spin
rotation of about $1.1$ deg can be observed when the duration of experiment is
$8 \tau_n$ and the guiding ring radius is $2$~mm. This is close to the
precision of $1$ deg achievable in a standard configuration.\footnote{Better
precision has been obtained in an experiment in
\cite{hasegawa_falsification_2012}.} Since the total rotation is proportional
to the duration of the experiment, the effect will be twice as large at about
$2.2$ deg for experiments that last twice as long, and even larger for longer
times. Note that at $8 \tau_n$ the duration of the experiment exceeds neutron
mean lifetime by eight times, which results in loss of signal. Although this is
a restriction it is not a critical issue as long as the resulting signal can be
detected.\footnote{Free neutron decay is described by $N(t) = N(0) e^{-t /
\tau_{n}}$, where $N(0)$ is the initial number of neutrons. This means after
time period $t = 2\tau_{n}$ about $13\%$ of the initial population remains,
while only $0.03 \%$ is left after $8\tau_{n}$ has passed. Whether or not the
final population of neutrons can be measured at the end of the experiment
depends on the detector sensitivity and the initial size of the neutron
population.}

For an example of how the same total rotation angle of $1.1$~deg can be
achieved in a different parameter regime, note that the total rotation angle
$\alpha_T$ is proportional to speed $v$ and experiment time $t$, and inversely
proportional to radius $r$ of the guide. Thus increasing speed to $2 \cdot
10^3$~$\mathrm{m s^{-1}}$ while reducing experiment time to $\tau_n$ also
yields the result $1.1$~deg. Because now the experiment time is equal to the
mean lifetime of neutrons the resulting signal would be significantly stronger.
On the other hand, larger speed means creating suitable beam width and
divergence is more challenging because the critical angle must be less than
$0.1$~deg in this regime.

In summary, we see that while in principle the parameters can be controlled,
which allows for different regimes of implementation, the latter also pose
different restrictions and challenges for realizing the experiment.

\subsection{Wave mechanics}\label{sec:wave_mechanics}

In the above description of the experiment, we used two different approaches to
describe the two different degrees of freedom of coordinate and spin. We
modeled spin evolution in terms of differential geometry; this allowed for an
efficient computation of Wigner rotation. On the other hand, we treated spatial
propagation using neutron optics. This is justified by the relatively large
wavelength of cold neutrons in comparison to interatomic distance. However, a
more detailed and accurate approach is to describe spatial propagation in terms
of quantum wave mechanics. Such an approach has been taken in
\cite{nesvizhevsky_etal_2008,nesvizhevsky_etal_2009,nesvizhevsky_etal_2010}. In
this section, we will briefly describe this work and then comment on the extent
to which the same approach can be adopted for our experiment. A more thorough
quantitative analysis remains to be carried out in the future.

In the wave mechanics picture, neutrons are described as matter waves that
propagate in a potential well formed by the repulsive Fermi potential of the
curved guide and the effective centrifugal potential
\cite{nesvizhevsky_etal_2009}. Whereas \cite{nesvizhevsky_etal_2009} assume the
neutron guide has the shape of a quarter (or a smaller segment) of a circle, we
assume a ring-like waveguide $C$. The Schroedinger equation describing the
situation remains the same as in \cite{nesvizhevsky_etal_2008}. The particle's
dynamics is described in cylindrical coordinates $(\rho, \phi)$ in the plane of
the figure, where $\rho$ is radial distance and $\phi$ is the angle. The states
of the neutrons moving parallel to the surface of the guide have angular
momentum which is close to the classical value. The latter justifies
approximating the angular coordinate $\phi$ as a classical continuous variable
and describing the propagation along $\phi$ using classical expression
\begin{align}\label{eq:neutron-classical-phi-1}
\phi = \frac{v t}{r}.
\end{align}
However, the radial dependence displays quantum behavior. The most significant
aspect of radial motion for the purposes of our experiment is the existence of
deeply bound states in the potential well formed by the effective centrifugal
potential and the Fermi potential of the guide. These sates are quasistationary
since there is a small probability they will tunnel through the potential
barrier. The radial dependence of neutron states in the triangular shaped
potential well in the vicinity of the curved guide can be found by solving the
Schroedinger equation for the potentials in question
\cite{nesvizhevsky_etal_2008}. Treating $\phi$ as classical variable, one
obtains an expression for the neutron flux describing the evolution of wave
packet in the vicinity of the curved guide. The existence of deeply bound
quasi-stationary states has been experimentally confirmed in
\cite{nesvizhevsky_etal_2009}.

The importance of this for our experiment is as follows. Whereas in
\cite{nesvizhevsky_etal_2008, nesvizhevsky_etal_2009} the curved mirror has the
shape of a segment of circle, in our setup the curved waveguide is a full
circle. The formal description in terms of the Schroedinger equation remains
the same and so we expect the results of such an analysis to yield states with
a structure similar to those described in the above mentioned works. However,
we note that the coordinate and spin will be still described by two different
formalisms: the spatial propagation in terms of Schroedinger equation and the
spin rotation using the differential geometric formalism. The two formalisms
complement each other. While the wave mechanics theory can be used to estimate
the flux of neutrons that leave the apparatus, the differential geometric
approach provides information about the total spin rotation angle.

\subsection{Challenges}

Although neutron experiments have achieved a high level of sophistication, the
experiment described above remains challenging because it demands parameter
regimes that seem to go beyond the current level of techniques. In this
section, we briefly discuss the difficulties that arise along with prospects of
mitigating the issues. We hope this will inspire further analysis and
technological developments.

The primary concern is the large number of revolutions that neutrons must
complete in order to achieve the observable $1$~deg of total spin rotation at
the end of the experiment. Depending on the implementation parameters, a single
revolution produces a minuscule spin rotation of $\sim\! 10^{-9} -
10^{-8}$~deg, which requires about $\sim\! 10^8 - 10^9$ revolutions to reach
$1$~deg of spin rotation. In the neutron optics approach this requires that the
neutron survive the large number of reflections in the circular waveguide;
neutrons that approach the guide at a larger angle than the critical angle
$\theta$ will be absorbed. In the wave mechanics picture this translates into
the question of how large is the flux of neutrons leaving the circular
waveguide after experiment time $t$ has passed.

The requirement that only neutrons with grazing incidence under the critical
angle $\theta$ contribute to the experiment severely restricts beam width.
Assuming for a moment that the beam is parallel, and the particles enter the
curved guide such that the top part of the beam is aligned with the tangent to
the perimeter of the curved guide, the critical angle limits the maximum width
of the beam to be $r [1 - \cos(\theta)]$.\footnote{We would like to thank an
anonymous reviewer for pointing this out.} For instance, if $\theta = 0.3$~deg
and $r = 2$~mm, the maximum width is $27$~nm. However, in reality the neutron
beam has angular divergence; typical divergence $\Delta\theta$ in neutron
experimentation facilities is from $0.5-1.0$~deg. In this case, the notion of
beam width is an approximation since the width increases in the direction of
propagation, and it is also useful to consider the angle of acceptance, i.e.\
the arc that spans the region where particles hit the curved guide. For the
parallel beam, the angle of acceptance is $\theta$, whereas for the divergent
beam it is larger, $\theta + \Delta\theta$. The reduction of signal due to the
narrow beam and small critical angle, coupled with long experiment time might
make neutron detection extremely challenging. A possible strategy for
increasing signal is to increase the critical angle by using neutrons with
lower speed. Another is to employ Bragg mirrors, or Ni-Ti super-mirrors which
are capable of achieving higher reflection angles.

A further concern is surface roughness. It can be viewed as transforming the
velocity component which is parallel to the guiding surface to a component
which is normal, thus effectively increasing absorption of neutrons, which
ultimately, again, leads to reduction of signal. While a detailed theory of
surface roughness can be found in \cite{nesvizhevsky_etal_2008} and in
references therein, we note that in the realization of the whispering gallery
experiment in \cite{nesvizhevsky_etal_2010}, mirror surface roughness of
$0.4$~nm was achieved, which was much smaller than the characteristic size of
the quantum states involved. We thus assume that surface quality of a guiding
ring with such roughness, or even better, can be realized in the experiment.

Lastly, one must take into account external magnetic fields from various
sources that lead to systematic errors. For example, the Earth's magnetic field
causes Larmor precession of neutron spin. More generally, a wide variety of
external sources ranging from nearby machinery to people would all introduce
distortions that lead to unwanted effects and errors in sensitive measurements.
In order to reduce such effects, the following scheme can be used. First, we
assume the experiment is carried out using magnetic shielding techniques. For
instance, in \cite{altarev_etal_2015} the authors report high quality magnetic
shielding where residual fields $B_R$ were observed to be lower than $1$~nT.
However, a magnetic field of $1$~nT would still cause spin precession at
several degrees per second, making observation of the significantly smaller
Wigner rotation impossible. To counter this, we adopt the following
compensation method. We apply a guide field $B_G$ that defines an axis
$\boldsymbol{\hat{n}}$ of spin rotation. Its magnitude must be large enough so
that the angular deviation $\beta$ from $\boldsymbol{\hat{n}}$ caused by the
residual field is negligible. For this to be true, $B_R / B_G = \tan\beta$
must be smaller than $10^{-3}$. This ensures that even in the worst case when
$B_R$ and $B_G$ are orthogonal, spin rotation axis is in practice defined by
the guiding field. Finally, we use two circular guide rings $C$, one where
neutrons orbit in the clockwise and the other where they move in the
counter-clockwise direction. The axis of the neutrons' circular movement must
coincide with the direction of the guide field $B_G$. This ensures that, in one
direction, the guide-field-induced Larmor precession and Wigner rotation add,
whereas in the other direction the Wigner rotation is subtracted from the
Larmor rotation. Measuring the final spin rotation of the two streams of
neutrons and noting their difference yields the Wigner rotation angle.
Incidentally, the scheme halves the required experiment time because the Wigner
rotation of $0.5$~deg in each direction results in the total difference of
$1$~deg.

In summary, we believe that by combining the above mentioned strategies
progress can be made towards realizing the experiment in the future.

\section{Conclusion}\label{sec:conclusion}

We have proposed an experiment that allows to verify Wigner's rotation for
massive spin-$1/2$ particles in the non-relativistic regime with cold neutrons.
The reason one can detect Wigner's rotation at these low velocities rests on
the cumulative character of the phenomenon: although each revolution around the
circle at $v = 10^3$~$\mathrm{m s^{-1}}$ produces a minuscule rotation of
$\sim\!\! 10^{-9}$~deg, this can be amplified to an observable $1$~deg by
letting the particle complete a sufficient number of revolutions. However, the
proposed experiment remains challenging at present since it requires parameter
regimes that go beyond current level of experimentation techniques. We hope
that steady progress in the field will make implementation of the experiment
possible in the future.

\section{Acknowledgments}

We would like to thank the anonymous reviewers for raising a number of
important points, as well as for helpful suggestions that improved the quality
of the paper. VP acknowledges partial support by the Estonian Research Council
(Eesti Teadusagentuur, ETAG) through grants PSG489 and PRG946. CP acknowledges
support by the excellence cluster QuantumFrontiers of the German Research
Foundation (Deutsche Forschungsgemeinschaft, DFG) under Germany's Excellence
Strategy, EXC-2123 QuantumFrontiers, 390837967; support by the German Research
Foundation (Deutsche Forschungsgemeinschaft, DFG), Project Number 420243324 and
by the Transilvania Fellowships for Visiting Professors grant 2024 of the
Transilvania University of Brasov.

\bibliography{experiment-wigner-2023}

\begin{thebibliography}{29}%
\makeatletter
\providecommand \@ifxundefined [1]{%
 \@ifx{#1\undefined}
}%
\providecommand \@ifnum [1]{%
 \ifnum #1\expandafter \@firstoftwo
 \else \expandafter \@secondoftwo
 \fi
}%
\providecommand \@ifx [1]{%
 \ifx #1\expandafter \@firstoftwo
 \else \expandafter \@secondoftwo
 \fi
}%
\providecommand \natexlab [1]{#1}%
\providecommand \enquote  [1]{``#1''}%
\providecommand \bibnamefont  [1]{#1}%
\providecommand \bibfnamefont [1]{#1}%
\providecommand \citenamefont [1]{#1}%
\providecommand \href@noop [0]{\@secondoftwo}%
\providecommand \href [0]{\begingroup \@sanitize@url \@href}%
\providecommand \@href[1]{\@@startlink{#1}\@@href}%
\providecommand \@@href[1]{\endgroup#1\@@endlink}%
\providecommand \@sanitize@url [0]{\catcode `\\12\catcode `\$12\catcode
  `\&12\catcode `\#12\catcode `\^12\catcode `\_12\catcode `\%12\relax}%
\providecommand \@@startlink[1]{}%
\providecommand \@@endlink[0]{}%
\providecommand \url  [0]{\begingroup\@sanitize@url \@url }%
\providecommand \@url [1]{\endgroup\@href {#1}{\urlprefix }}%
\providecommand \urlprefix  [0]{URL }%
\providecommand \Eprint [0]{\href }%
\providecommand \doibase [0]{https://doi.org/}%
\providecommand \selectlanguage [0]{\@gobble}%
\providecommand \bibinfo  [0]{\@secondoftwo}%
\providecommand \bibfield  [0]{\@secondoftwo}%
\providecommand \translation [1]{[#1]}%
\providecommand \BibitemOpen [0]{}%
\providecommand \bibitemStop [0]{}%
\providecommand \bibitemNoStop [0]{.\EOS\space}%
\providecommand \EOS [0]{\spacefactor3000\relax}%
\providecommand \BibitemShut  [1]{\csname bibitem#1\endcsname}%
\let\auto@bib@innerbib\@empty
\bibitem [{\citenamefont {Rauch}\ and\ \citenamefont
  {Werner}(2000)}]{rauch_werner_2000}%
  \BibitemOpen
  \bibfield  {author} {\bibinfo {author} {\bibfnamefont {H.}~\bibnamefont
  {Rauch}}\ and\ \bibinfo {author} {\bibfnamefont {S.~A.}\ \bibnamefont
  {Werner}},\ }\href@noop {} {\emph {\bibinfo {title} {Neutron
  {I}nterferometry}}}\ (\bibinfo  {publisher} {Clarendon Press},\ \bibinfo
  {address} {Oxford},\ \bibinfo {year} {2000})\BibitemShut {NoStop}%
\bibitem [{\citenamefont {Sponar}\ \emph
  {et~al.}(2010{\natexlab{a}})\citenamefont {Sponar}, \citenamefont {Klepp},
  \citenamefont {Loidl}, \citenamefont {Filipp}, \citenamefont
  {Durstberger-Rennhofer}, \citenamefont {Bertlmann}, \citenamefont {Badurek},
  \citenamefont {Rauch},\ and\ \citenamefont {Hasegawa}}]{sponar_etal_2010}%
  \BibitemOpen
  \bibfield  {author} {\bibinfo {author} {\bibfnamefont {S.}~\bibnamefont
  {Sponar}}, \bibinfo {author} {\bibfnamefont {J.}~\bibnamefont {Klepp}},
  \bibinfo {author} {\bibfnamefont {R.}~\bibnamefont {Loidl}}, \bibinfo
  {author} {\bibfnamefont {S.}~\bibnamefont {Filipp}}, \bibinfo {author}
  {\bibfnamefont {K.}~\bibnamefont {Durstberger-Rennhofer}}, \bibinfo {author}
  {\bibfnamefont {R.~A.}\ \bibnamefont {Bertlmann}}, \bibinfo {author}
  {\bibfnamefont {G.}~\bibnamefont {Badurek}}, \bibinfo {author} {\bibfnamefont
  {H.}~\bibnamefont {Rauch}},\ and\ \bibinfo {author} {\bibfnamefont
  {Y.}~\bibnamefont {Hasegawa}},\ }\href
  {https://doi.org/10.1103/PhysRevA.81.042113} {\bibfield  {journal} {\bibinfo
  {journal} {Phys. Rev. A}\ }\textbf {\bibinfo {volume} {81}},\ \bibinfo
  {pages} {042113} (\bibinfo {year} {2010}{\natexlab{a}})}\BibitemShut
  {NoStop}%
\bibitem [{\citenamefont {Hasegawa}\ \emph {et~al.}(2006)\citenamefont
  {Hasegawa}, \citenamefont {Loidl}, \citenamefont {Badurek}, \citenamefont
  {Baron},\ and\ \citenamefont {Rauch}}]{hasegawa_etal_2006}%
  \BibitemOpen
  \bibfield  {author} {\bibinfo {author} {\bibfnamefont {Y.}~\bibnamefont
  {Hasegawa}}, \bibinfo {author} {\bibfnamefont {R.}~\bibnamefont {Loidl}},
  \bibinfo {author} {\bibfnamefont {G.}~\bibnamefont {Badurek}}, \bibinfo
  {author} {\bibfnamefont {M.}~\bibnamefont {Baron}},\ and\ \bibinfo {author}
  {\bibfnamefont {H.}~\bibnamefont {Rauch}},\ }\href
  {https://doi.org/10.1103/PhysRevLett.97.230401} {\bibfield  {journal}
  {\bibinfo  {journal} {Phys. Rev. Lett.}\ }\textbf {\bibinfo {volume} {97}},\
  \bibinfo {pages} {230401} (\bibinfo {year} {2006})}\BibitemShut {NoStop}%
\bibitem [{\citenamefont {Bartosik}\ \emph {et~al.}(2009)\citenamefont
  {Bartosik}, \citenamefont {Klepp}, \citenamefont {Schmitzer}, \citenamefont
  {Sponar}, \citenamefont {Cabello}, \citenamefont {Rauch},\ and\ \citenamefont
  {Hasegawa}}]{bartosik_etal_2009}%
  \BibitemOpen
  \bibfield  {author} {\bibinfo {author} {\bibfnamefont {H.}~\bibnamefont
  {Bartosik}}, \bibinfo {author} {\bibfnamefont {J.}~\bibnamefont {Klepp}},
  \bibinfo {author} {\bibfnamefont {C.}~\bibnamefont {Schmitzer}}, \bibinfo
  {author} {\bibfnamefont {S.}~\bibnamefont {Sponar}}, \bibinfo {author}
  {\bibfnamefont {A.}~\bibnamefont {Cabello}}, \bibinfo {author} {\bibfnamefont
  {H.}~\bibnamefont {Rauch}},\ and\ \bibinfo {author} {\bibfnamefont
  {Y.}~\bibnamefont {Hasegawa}},\ }\href
  {https://doi.org/10.1103/PhysRevLett.103.040403} {\bibfield  {journal}
  {\bibinfo  {journal} {Phys. Rev. Lett.}\ }\textbf {\bibinfo {volume} {103}},\
  \bibinfo {pages} {040403} (\bibinfo {year} {2009})}\BibitemShut {NoStop}%
\bibitem [{\citenamefont {Klepp}\ \emph {et~al.}(2014)\citenamefont {Klepp},
  \citenamefont {Sponar},\ and\ \citenamefont {Hasegawa}}]{klepp_etal_2014}%
  \BibitemOpen
  \bibfield  {author} {\bibinfo {author} {\bibfnamefont {J.}~\bibnamefont
  {Klepp}}, \bibinfo {author} {\bibfnamefont {S.}~\bibnamefont {Sponar}},\ and\
  \bibinfo {author} {\bibfnamefont {Y.}~\bibnamefont {Hasegawa}},\ }\href
  {https://doi.org/10.1093/ptep/ptu085} {\bibfield  {journal} {\bibinfo
  {journal} {Progress of Theoretical and Experimental Physics}\ }\textbf
  {\bibinfo {volume} {2014}},\ \bibinfo {pages} {082A01} (\bibinfo {year}
  {2014})}\BibitemShut {NoStop}%
\bibitem [{\citenamefont {Sponar}\ \emph {et~al.}(2021)\citenamefont {Sponar},
  \citenamefont {Sedmik}, \citenamefont {Pitschmann}, \citenamefont {Abele},\
  and\ \citenamefont {Hasegawa}}]{sponar_2021}%
  \BibitemOpen
  \bibfield  {author} {\bibinfo {author} {\bibfnamefont {S.}~\bibnamefont
  {Sponar}}, \bibinfo {author} {\bibfnamefont {R.~I.~P.}\ \bibnamefont
  {Sedmik}}, \bibinfo {author} {\bibfnamefont {M.}~\bibnamefont {Pitschmann}},
  \bibinfo {author} {\bibfnamefont {H.}~\bibnamefont {Abele}},\ and\ \bibinfo
  {author} {\bibfnamefont {Y.}~\bibnamefont {Hasegawa}},\ }\href
  {https://doi.org/10.1038/s42254-021-00298-2} {\bibfield  {journal} {\bibinfo
  {journal} {Nature Reviews Physics}\ }\textbf {\bibinfo {volume} {3}},\
  \bibinfo {pages} {309–327} (\bibinfo {year} {2021})}\BibitemShut {NoStop}%
\bibitem [{\citenamefont {Thomas}(1926)}]{thomas_motion_1926}%
  \BibitemOpen
  \bibfield  {author} {\bibinfo {author} {\bibfnamefont {L.~H.}\ \bibnamefont
  {Thomas}},\ }\href {https://doi.org/10.1038/117514a0} {\bibfield  {journal}
  {\bibinfo  {journal} {Nature}\ }\textbf {\bibinfo {volume} {117}},\ \bibinfo
  {pages} {514–514} (\bibinfo {year} {1926})}\BibitemShut {NoStop}%
\bibitem [{\citenamefont {Gingrich}\ and\ \citenamefont
  {Adami}(2002)}]{gingrich_quantum_2002}%
  \BibitemOpen
  \bibfield  {author} {\bibinfo {author} {\bibfnamefont {R.~M.}\ \bibnamefont
  {Gingrich}}\ and\ \bibinfo {author} {\bibfnamefont {C.}~\bibnamefont
  {Adami}},\ }\href {https://doi.org/10.1103/PhysRevLett.89.270402} {\bibfield
  {journal} {\bibinfo  {journal} {Physical Review Letters}\ }\textbf {\bibinfo
  {volume} {89}},\ \bibinfo {pages} {270402} (\bibinfo {year}
  {2002})}\BibitemShut {NoStop}%
\bibitem [{\citenamefont {Peres}\ \emph {et~al.}(2002)\citenamefont {Peres},
  \citenamefont {Scudo},\ and\ \citenamefont {Terno}}]{peres_quantum_2002}%
  \BibitemOpen
  \bibfield  {author} {\bibinfo {author} {\bibfnamefont {A.}~\bibnamefont
  {Peres}}, \bibinfo {author} {\bibfnamefont {P.~F.}\ \bibnamefont {Scudo}},\
  and\ \bibinfo {author} {\bibfnamefont {D.~R.}\ \bibnamefont {Terno}},\ }\href
  {https://doi.org/10.1103/PhysRevLett.88.230402} {\bibfield  {journal}
  {\bibinfo  {journal} {Physical Review Letters}\ }\textbf {\bibinfo {volume}
  {88}},\ \bibinfo {pages} {230402} (\bibinfo {year} {2002})}\BibitemShut
  {NoStop}%
\bibitem [{\citenamefont {Friis}\ \emph {et~al.}(2010)\citenamefont {Friis},
  \citenamefont {Bertlmann},\ and\ \citenamefont
  {Huber}}]{friis_relativistic_2010}%
  \BibitemOpen
  \bibfield  {author} {\bibinfo {author} {\bibfnamefont {N.}~\bibnamefont
  {Friis}}, \bibinfo {author} {\bibfnamefont {R.~A.}\ \bibnamefont
  {Bertlmann}},\ and\ \bibinfo {author} {\bibfnamefont {M.}~\bibnamefont
  {Huber}},\ }\href {https://doi.org/10.1103/PhysRevA.81.042114} {\bibfield
  {journal} {\bibinfo  {journal} {Physical Review A}\ }\textbf {\bibinfo
  {volume} {81}},\ \bibinfo {pages} {042114} (\bibinfo {year}
  {2010})}\BibitemShut {NoStop}%
\bibitem [{\citenamefont {Palge}\ and\ \citenamefont
  {Dunningham}(2015)}]{palge_behavior_2015}%
  \BibitemOpen
  \bibfield  {author} {\bibinfo {author} {\bibfnamefont {V.}~\bibnamefont
  {Palge}}\ and\ \bibinfo {author} {\bibfnamefont {J.}~\bibnamefont
  {Dunningham}},\ }\href {https://doi.org/10.1016/j.aop.2015.09.028} {\bibfield
   {journal} {\bibinfo  {journal} {Annals of Physics}\ }\textbf {\bibinfo
  {volume} {363}},\ \bibinfo {pages} {275–304} (\bibinfo {year}
  {2015})}\BibitemShut {NoStop}%
\bibitem [{\citenamefont {Barr}\ \emph {et~al.}(2023)\citenamefont {Barr},
  \citenamefont {Caban},\ and\ \citenamefont {Rembieliński}}]{barr_etal_2023}%
  \BibitemOpen
  \bibfield  {author} {\bibinfo {author} {\bibfnamefont {A.~J.}\ \bibnamefont
  {Barr}}, \bibinfo {author} {\bibfnamefont {P.}~\bibnamefont {Caban}},\ and\
  \bibinfo {author} {\bibfnamefont {J.}~\bibnamefont {Rembieliński}},\ }\href
  {https://doi.org/10.22331/q-2023-07-27-1070} {\bibfield  {journal} {\bibinfo
  {journal} {Quantum}\ }\textbf {\bibinfo {volume} {7}},\ \bibinfo {pages}
  {1070} (\bibinfo {year} {2023})},\ \Eprint {https://arxiv.org/abs/2204.11063}
  {arXiv:2204.11063 [quant-ph]} \BibitemShut {NoStop}%
\bibitem [{\citenamefont {Palge}\ and\ \citenamefont
  {Pfeifer}(2024)}]{palge_pfeifer_2024}%
  \BibitemOpen
  \bibfield  {author} {\bibinfo {author} {\bibfnamefont {V.}~\bibnamefont
  {Palge}}\ and\ \bibinfo {author} {\bibfnamefont {C.}~\bibnamefont
  {Pfeifer}},\ }\href {https://doi.org/10.1103/PhysRevA.109.032206} {\bibfield
  {journal} {\bibinfo  {journal} {Physical Review A}\ }\textbf {\bibinfo
  {volume} {109}},\ \bibinfo {pages} {032206} (\bibinfo {year} {2024})},\
  \Eprint {https://arxiv.org/abs/2310.08121} {arXiv:2310.08121 [quant-ph]}
  \BibitemShut {NoStop}%
\bibitem [{\citenamefont {Halpern}(1968)}]{halpern_special_1968}%
  \BibitemOpen
  \bibfield  {author} {\bibinfo {author} {\bibfnamefont {F.~R.}\ \bibnamefont
  {Halpern}},\ }\href@noop {} {\emph {\bibinfo {title} {Special {Relativity}
  and {Quantum} {Mechanics}}}}\ (\bibinfo  {publisher} {Prentice-Hall},\
  \bibinfo {year} {1968})\BibitemShut {NoStop}%
\bibitem [{\citenamefont {Costella}\ \emph {et~al.}(2001)\citenamefont
  {Costella}, \citenamefont {McKellar},\ and\ \citenamefont
  {Rawlinson}}]{costella_2001}%
  \BibitemOpen
  \bibfield  {author} {\bibinfo {author} {\bibfnamefont {J.~P.}\ \bibnamefont
  {Costella}}, \bibinfo {author} {\bibfnamefont {B.~H.~J.}\ \bibnamefont
  {McKellar}},\ and\ \bibinfo {author} {\bibfnamefont {A.~A.}\ \bibnamefont
  {Rawlinson}},\ }\bibfield  {journal} {\bibinfo  {journal} {American Journal
  of Physics}\ }\textbf {\bibinfo {volume} {69}},\ \href
  {https://doi.org/10.1119/1.1371010} {10.1119/1.1371010} (\bibinfo {year}
  {2001})\BibitemShut {NoStop}%
\bibitem [{\citenamefont {Rhodes}\ and\ \citenamefont
  {Semon}(2004)}]{rhodes_relativistic_2004}%
  \BibitemOpen
  \bibfield  {author} {\bibinfo {author} {\bibfnamefont {J.~A.}\ \bibnamefont
  {Rhodes}}\ and\ \bibinfo {author} {\bibfnamefont {M.~D.}\ \bibnamefont
  {Semon}},\ }\href {https://doi.org/10.1119/1.1652040} {\bibfield  {journal}
  {\bibinfo  {journal} {American Journal of Physics}\ }\textbf {\bibinfo
  {volume} {72}},\ \bibinfo {pages} {943–960} (\bibinfo {year}
  {2004})}\BibitemShut {NoStop}%
\bibitem [{\citenamefont {Wilczek}\ and\ \citenamefont
  {Shapere}(1989)}]{wilczek_shapere_1989}%
  \BibitemOpen
  \bibfield  {author} {\bibinfo {author} {\bibfnamefont {F.}~\bibnamefont
  {Wilczek}}\ and\ \bibinfo {author} {\bibfnamefont {A.}~\bibnamefont
  {Shapere}},\ }\href {https://doi.org/10.1142/0613} {\emph {\bibinfo {title}
  {Geometric Phases in Physics}}},\ Advanced Series in Mathematical Physics:
  Volume 5\ (\bibinfo  {publisher} {World Scientific},\ \bibinfo {year}
  {1989})\BibitemShut {NoStop}%
\bibitem [{\citenamefont {Lyre}(2014)}]{lyre_2014}%
  \BibitemOpen
  \bibfield  {author} {\bibinfo {author} {\bibfnamefont {H.}~\bibnamefont
  {Lyre}},\ }\href {https://doi.org/10.1016/j.shpsb.2014.08.013} {\bibfield
  {journal} {\bibinfo  {journal} {Studies in History and Philosophy of Science
  Part B: Studies in History and Philosophy of Modern Physics}\ }\textbf
  {\bibinfo {volume} {48}},\ \bibinfo {pages} {45–51} (\bibinfo {year}
  {2014})}\BibitemShut {NoStop}%
\bibitem [{\citenamefont {Simon}(1983)}]{simon_1983_holonomy}%
  \BibitemOpen
  \bibfield  {author} {\bibinfo {author} {\bibfnamefont {B.}~\bibnamefont
  {Simon}},\ }\href@noop {} {\bibfield  {journal} {\bibinfo  {journal}
  {Physical Review Letters}\ }\textbf {\bibinfo {volume} {51}},\ \bibinfo
  {pages} {2167} (\bibinfo {year} {1983})}\BibitemShut {NoStop}%
\bibitem [{\citenamefont {Cisowski}\ \emph {et~al.}(2022)\citenamefont
  {Cisowski}, \citenamefont {Götte},\ and\ \citenamefont
  {Franke-Arnold}}]{cisowski_2022}%
  \BibitemOpen
  \bibfield  {author} {\bibinfo {author} {\bibfnamefont {C.}~\bibnamefont
  {Cisowski}}, \bibinfo {author} {\bibfnamefont {J.}~\bibnamefont {Götte}},\
  and\ \bibinfo {author} {\bibfnamefont {S.}~\bibnamefont {Franke-Arnold}},\
  }\href@noop {} {\bibfield  {journal} {\bibinfo  {journal} {Reviews of Modern
  Physics}\ }\textbf {\bibinfo {volume} {94}},\ \bibinfo {pages} {031001}
  (\bibinfo {year} {2022})}\BibitemShut {NoStop}%
\bibitem [{\citenamefont {Hasegawa}\ and\ \citenamefont
  {Badurek}(1999)}]{hasegawa_noncommuting_1999}%
  \BibitemOpen
  \bibfield  {author} {\bibinfo {author} {\bibfnamefont {Y.}~\bibnamefont
  {Hasegawa}}\ and\ \bibinfo {author} {\bibfnamefont {G.}~\bibnamefont
  {Badurek}},\ }\href {https://doi.org/10.1103/PhysRevA.59.4614} {\bibfield
  {journal} {\bibinfo  {journal} {Physical Review A}\ }\textbf {\bibinfo
  {volume} {59}},\ \bibinfo {pages} {4614–4622} (\bibinfo {year}
  {1999})}\BibitemShut {NoStop}%
\bibitem [{\citenamefont {Sponar}\ \emph
  {et~al.}(2010{\natexlab{b}})\citenamefont {Sponar}, \citenamefont {Klepp},
  \citenamefont {Zeiner}, \citenamefont {Badurek},\ and\ \citenamefont
  {Hasegawa}}]{sponar_violation_2010}%
  \BibitemOpen
  \bibfield  {author} {\bibinfo {author} {\bibfnamefont {S.}~\bibnamefont
  {Sponar}}, \bibinfo {author} {\bibfnamefont {J.}~\bibnamefont {Klepp}},
  \bibinfo {author} {\bibfnamefont {C.}~\bibnamefont {Zeiner}}, \bibinfo
  {author} {\bibfnamefont {G.}~\bibnamefont {Badurek}},\ and\ \bibinfo {author}
  {\bibfnamefont {Y.}~\bibnamefont {Hasegawa}},\ }\href
  {https://doi.org/10.1016/j.physleta.2009.11.017} {\bibfield  {journal}
  {\bibinfo  {journal} {Physics Letters A}\ }\textbf {\bibinfo {volume}
  {374}},\ \bibinfo {pages} {431–434} (\bibinfo {year}
  {2010}{\natexlab{b}})}\BibitemShut {NoStop}%
\bibitem [{\citenamefont {Demirel}\ \emph {et~al.}(2015)\citenamefont
  {Demirel}, \citenamefont {Sponar},\ and\ \citenamefont
  {Hasegawa}}]{demirel_measurement_2015}%
  \BibitemOpen
  \bibfield  {author} {\bibinfo {author} {\bibfnamefont {B.}~\bibnamefont
  {Demirel}}, \bibinfo {author} {\bibfnamefont {S.}~\bibnamefont {Sponar}},\
  and\ \bibinfo {author} {\bibfnamefont {Y.}~\bibnamefont {Hasegawa}},\ }\href
  {https://doi.org/10.1088/1367-2630/17/2/023065} {\bibfield  {journal}
  {\bibinfo  {journal} {New Journal of Physics}\ }\textbf {\bibinfo {volume}
  {17}},\ \bibinfo {pages} {023065} (\bibinfo {year} {2015})}\BibitemShut
  {NoStop}%
\bibitem [{\citenamefont {Boothroyd}(2020)}]{boothroyd_2020}%
  \BibitemOpen
  \bibfield  {author} {\bibinfo {author} {\bibfnamefont {A.~T.}\ \bibnamefont
  {Boothroyd}},\ }\href@noop {} {\emph {\bibinfo {title} {Principles of neutron
  scattering from condensed matter}}}\ (\bibinfo  {publisher} {Oxford
  University Press},\ \bibinfo {year} {2020})\BibitemShut {NoStop}%
\bibitem [{\citenamefont {Hasegawa}\ \emph {et~al.}(2012)\citenamefont
  {Hasegawa}, \citenamefont {Schmitzer}, \citenamefont {Bartosik},
  \citenamefont {Klepp}, \citenamefont {Sponar}, \citenamefont
  {Durstberger-Rennhofer},\ and\ \citenamefont
  {Badurek}}]{hasegawa_falsification_2012}%
  \BibitemOpen
  \bibfield  {author} {\bibinfo {author} {\bibfnamefont {Y.}~\bibnamefont
  {Hasegawa}}, \bibinfo {author} {\bibfnamefont {C.}~\bibnamefont {Schmitzer}},
  \bibinfo {author} {\bibfnamefont {H.}~\bibnamefont {Bartosik}}, \bibinfo
  {author} {\bibfnamefont {J.}~\bibnamefont {Klepp}}, \bibinfo {author}
  {\bibfnamefont {S.}~\bibnamefont {Sponar}}, \bibinfo {author} {\bibfnamefont
  {K.}~\bibnamefont {Durstberger-Rennhofer}},\ and\ \bibinfo {author}
  {\bibfnamefont {G.}~\bibnamefont {Badurek}},\ }\href
  {https://doi.org/10.1088/1367-2630/14/2/023039} {\bibfield  {journal}
  {\bibinfo  {journal} {New Journal of Physics}\ }\textbf {\bibinfo {volume}
  {14}},\ \bibinfo {pages} {023039} (\bibinfo {year} {2012})}\BibitemShut
  {NoStop}%
\bibitem [{\citenamefont {Nesvizhevsky}\ \emph {et~al.}(2008)\citenamefont
  {Nesvizhevsky}, \citenamefont {Petukhov}, \citenamefont {Protasov},\ and\
  \citenamefont {Voronin}}]{nesvizhevsky_etal_2008}%
  \BibitemOpen
  \bibfield  {author} {\bibinfo {author} {\bibfnamefont {V.~V.}\ \bibnamefont
  {Nesvizhevsky}}, \bibinfo {author} {\bibfnamefont {A.~K.}\ \bibnamefont
  {Petukhov}}, \bibinfo {author} {\bibfnamefont {K.~V.}\ \bibnamefont
  {Protasov}},\ and\ \bibinfo {author} {\bibfnamefont {A.~Y.}\ \bibnamefont
  {Voronin}},\ }\href@noop {} {\bibfield  {journal} {\bibinfo  {journal}
  {Physical Review A}\ }\textbf {\bibinfo {volume} {78}},\ \bibinfo {pages}
  {033616} (\bibinfo {year} {2008})}\BibitemShut {NoStop}%
\bibitem [{\citenamefont {Nesvizhevsky}\ \emph
  {et~al.}(2010{\natexlab{a}})\citenamefont {Nesvizhevsky}, \citenamefont
  {Voronin}, \citenamefont {Cubitt},\ and\ \citenamefont
  {Protasov}}]{nesvizhevsky_etal_2009}%
  \BibitemOpen
  \bibfield  {author} {\bibinfo {author} {\bibfnamefont {V.~V.}\ \bibnamefont
  {Nesvizhevsky}}, \bibinfo {author} {\bibfnamefont {A.~Y.}\ \bibnamefont
  {Voronin}}, \bibinfo {author} {\bibfnamefont {R.}~\bibnamefont {Cubitt}},\
  and\ \bibinfo {author} {\bibfnamefont {K.~V.}\ \bibnamefont {Protasov}},\
  }\href {https://doi.org/10.1038/NPHYS1478} {\bibfield  {journal} {\bibinfo
  {journal} {Nature Physics}\ }\textbf {\bibinfo {volume} {6}},\ \bibinfo
  {pages} {114–117} (\bibinfo {year} {2010}{\natexlab{a}})}\BibitemShut
  {NoStop}%
\bibitem [{\citenamefont {Nesvizhevsky}\ \emph
  {et~al.}(2010{\natexlab{b}})\citenamefont {Nesvizhevsky}, \citenamefont
  {Cubitt}, \citenamefont {Protasov},\ and\ \citenamefont
  {Voronin}}]{nesvizhevsky_etal_2010}%
  \BibitemOpen
  \bibfield  {author} {\bibinfo {author} {\bibfnamefont {V.~V.}\ \bibnamefont
  {Nesvizhevsky}}, \bibinfo {author} {\bibfnamefont {R.}~\bibnamefont
  {Cubitt}}, \bibinfo {author} {\bibfnamefont {K.~V.}\ \bibnamefont
  {Protasov}},\ and\ \bibinfo {author} {\bibfnamefont {A.~Y.}\ \bibnamefont
  {Voronin}},\ }\href@noop {} {\bibfield  {journal} {\bibinfo  {journal} {New
  Journal of Physics}\ }\textbf {\bibinfo {volume} {12}},\ \bibinfo {pages}
  {113050} (\bibinfo {year} {2010}{\natexlab{b}})}\BibitemShut {NoStop}%
\bibitem [{\citenamefont {Altarev}\ \emph {et~al.}(2015)\citenamefont
  {Altarev}, \citenamefont {Bales}, \citenamefont {Beck}, \citenamefont
  {Chupp}, \citenamefont {Fierlinger}, \citenamefont {Fierlinger},
  \citenamefont {Kuchler}, \citenamefont {Lins}, \citenamefont {Marino},
  \citenamefont {Niessen} \emph {et~al.}}]{altarev_etal_2015}%
  \BibitemOpen
  \bibfield  {author} {\bibinfo {author} {\bibfnamefont {I.}~\bibnamefont
  {Altarev}}, \bibinfo {author} {\bibfnamefont {M.}~\bibnamefont {Bales}},
  \bibinfo {author} {\bibfnamefont {D.}~\bibnamefont {Beck}}, \bibinfo {author}
  {\bibfnamefont {T.}~\bibnamefont {Chupp}}, \bibinfo {author} {\bibfnamefont
  {K.}~\bibnamefont {Fierlinger}}, \bibinfo {author} {\bibfnamefont
  {P.}~\bibnamefont {Fierlinger}}, \bibinfo {author} {\bibfnamefont
  {F.}~\bibnamefont {Kuchler}}, \bibinfo {author} {\bibfnamefont
  {T.}~\bibnamefont {Lins}}, \bibinfo {author} {\bibfnamefont {M.}~\bibnamefont
  {Marino}}, \bibinfo {author} {\bibfnamefont {B.}~\bibnamefont {Niessen}},
  \emph {et~al.},\ }\href@noop {} {\bibfield  {journal} {\bibinfo  {journal}
  {Journal of Applied Physics}\ }\textbf {\bibinfo {volume} {117}} (\bibinfo
  {year} {2015})}\BibitemShut {NoStop}%
\end{thebibliography}%

\end{document}